\documentclass[12pt]{article}
\usepackage{epsfig}
\newcommand{\Li}{\mathrm{Li}_2}
\newcommand{\Ri}{R(\infty)}
\newcommand{\dd}{{\mathrm d}}

\def\la{\mathrel{\mathpalette\fun <}}

\def\fun#1#2{\lower3.6pt\vbox{\baselineskip0pt\lineskip.9pt
  \ialign{$\mathsurround=0pt#1\hfil##\hfil$\crcr#2\crcr\sim\crcr}}}
\def\ba{\begin{eqnarray}}
\def\ea{\end{eqnarray}}
\def\order#1{{\mathcal O}\left(#1\right)}

\title{Higher order pair corrections to electron-positron annihilation}

\author{
A.B. Arbuzov\thanks{The present address: 
Bogoliubov Laboratory of Theoretical Physics, JINR, Dubna, Russia.}
}

\date{}

\begin{document}

\maketitle

\begin{itemize}
\item[$ $]
        {\em Dipartimento di Fisica Teorica, Universit\`a di Torino; \\
             INFN, Sezione di Torino, \\
             via Giuria 1, I-10125 Torino, Italy \/} 
\end{itemize}

\begin{abstract}
Radiative corrections due to pair production in process
of electron--positron annihilation are considered.
The main attention is paid to the corrections of the third
order. Simultaneous emission of photons and pairs
is taken into account. The leading log approximation and 
convolution procedure are used. Exponentiation is discussed. 
Theoretical uncertainty in description of higher order 
secondary pairs is estimated. 
\end{abstract}

\section{Introduction}

The high statistics and accuracy of experimental measurements
at LEP and SLC result in the outstanding level of precision in 
definition of the Standard Model parameters 
(see Ref.~\cite{Bardin:1999gt,Kobel:2000aw}
and references therein). In this situation effects of radiative
corrections are of great importance. 

One of particular contributions to the process of electron--positron
annihilation is the radiation of secondary pairs. 
In comparison with the photon radiation, it is relatively small, 
because it appears only starting from the ${\mathcal O}(\alpha^2)$ order. 
Nevertheless, the total effect of pair production can reach several permil 
(or even e few percent in particular cases) and
should be taken into account. We will consider here the reactions of the
following type:
\ba
e^+\ +\ e^-\ \to\ f\ +\ \bar{f}\ +\ (f'\bar{f}'(\gamma)),
\ea
where $f\bar{f}$ is the so-called primary fermionic pair, $f'\bar{f}'$ 
is the secondary one, and $(\gamma)$ stands for the usual photon radiation, 
which can take place in our case starting from the third order $\order{\alpha^2}$.
The primary pair is the one registered in the particular measurement.
The secondary one can be either real or virtual, and has to be accounted
as a radiative correction according to certain experimental conditions
(with some restrictions (cuts), if required). In principle, the pairs can be
of the same flavour, like in the process $e^+e^-\to2\mu^+2\mu^-$. When both
pairs are seen in the detector, one has to apply an algorithm to choose the 
primary pair.

The complete second order QED calculation for production of the 
initial state secondary pairs ($e^+e^-$, $\mu^+\mu^-$, and hadrons)
was performed in Refs.~\cite{Berends:1987ab,Kniehl:1988id}. 
The corresponding final state effect was considered in Ref.~\cite{Hoang:1995ht}. 

The interference between amplitudes for the initial and final state 
pair production has not been calculated systematically. An estimate of this 
effect will be presented in the Conclusions.

It is observed, that the ${\mathcal O}(\alpha^2)$ approximation is not 
adequate to provide the desirable precision, especially in the region close 
to the $Z$-peak. From another side, at LEP2, when the radiative return
is allowed, QED radiative corrections reach several hundred percent
and should be evaluated with care.
In order to achieve the precision, required by the experiments 
(see Ref.~\cite{Kobel:2000aw}), one should 
consider higher orders of the perturbation theory. 
Another important point is
that in the $Z$-peak region the sub--leading corrections 
are comparable in size with the leading ones. So, one has to go beyond
the leading log approximation (LLA).

The goal of the present paper is to describe and discuss the treatment 
of pair corrections, implemented in the semi--analytical code 
{\tt ZFITTER v.6.30}~\cite{Bardin:1999yd}. The main attention will 
be paid to the third order corrections. Some numerical illustrations
will be given. The uncertainty of the approach will be analyzed in the
Conclusions.

\section{Pair corrections in higher orders}

Starting from the ${\mathcal O}(\alpha^3)$ order, one has
amplitudes with simultaneous emission of pairs and photons. We treat
them as a part of pair corrections, because they have not been taken
into account in the traditional photonic RC. The main subject of this
article is to analyze this effect.

As usually in $e^+e^-$ annihilation, the most important contributions are 
from the initial state (IS) radiation (real and virtual).
In a typical high--energy experiment, the conditions of the 
Kinoshita--Lee--Nauenberg theorem are fulfilled for the final state (FS)
radiation. That is why the large logarithms (mass singularities) cancel
out in the FS correction, and the latter becomes relatively small.
From the other hand, a specific energy dependence of the kernel 
cross section can provoke an enhancement of the IS correction, like
it happens at the $Z$-peak or in the radiative return to the resonance 
at LEP2.

Pair corrections to the process of $e^+e^-\to f\bar{f}$ become ambiguous,
when one observes $4$-fermion final states. Nevertheless, the bulk of
the pair RC to the $2$-fermion observables is coming from the kinematical
situations, when the secondary pair is either virtual or escapes
registration (in the beam pipe or because of a small energy). On could try
to subtract from the experimental data all real secondary pairs by means
of a Monte Carlo generator. But that would destroy the cancellation
between the real and virtual pairs, and the corrections would become even
more pronounced and dangerous. One can get a serious numerical instability,
trying to subtract both the virtual and the real pairs by means of different
codes. Even if the subtraction of the pure second order pairs is evaluated
perfectly, we have to keep in mind the higher order corrections due to 
instantaneous emission of pairs and photons. The cancellation between 
the virtual and real pair RC is important not only in the $\order{\alpha^2}$,
but in higher orders as well. Indeed, we would end up with a completely 
wrong third order corrections (the error would exceed the entire effect),
if we were using for the subtraction of virtual and real pair contributions
two different codes, one with the third order pair-photon effect and 
another one without it.

%--------------------------Figure~1-----------------------
\begin{figure}[ht]
\begin{center}
\epsfig{file=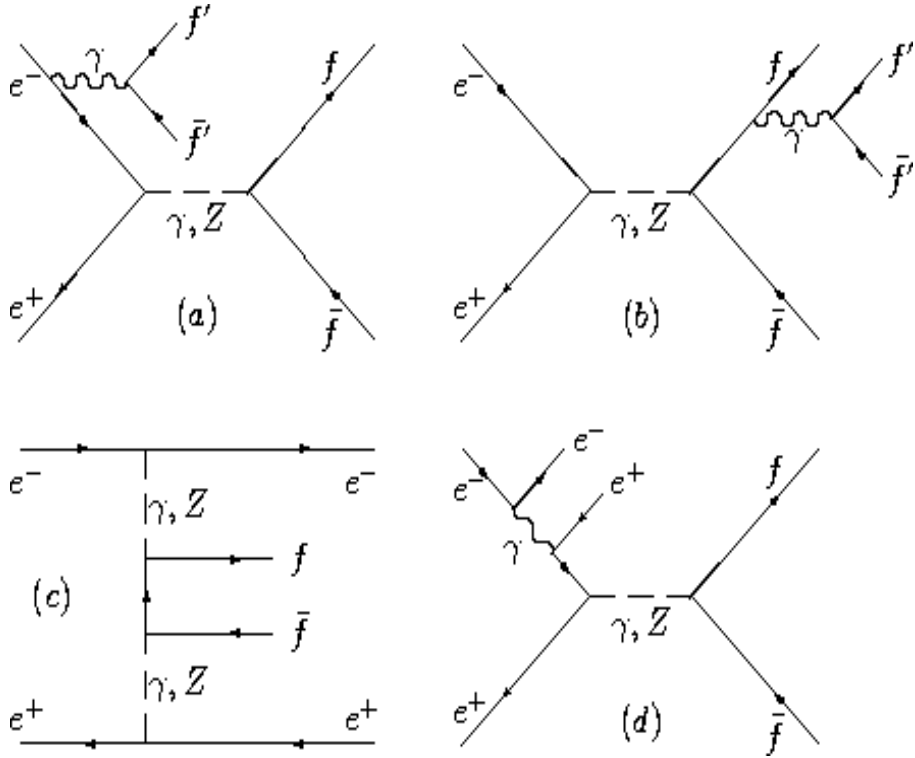,width=12cm,height=10cm}
\end{center}
\caption{Different types of Feynman diagrams for real pair production.}
\label{figure:1}
\end{figure}
%--------------------------Figure~1-----------------------

Here we will follow the definition of signal and background according to
Feynman diagrams, as discussed in Ref.~\cite{Kobel:2000aw}. In Figure~1
we show several representatives of Feynman graphs for real pair production 
processes. Only the diagrams Fig.~1$a$ and Fig.~1$b$ (plus the ones with pair 
production via a virtual photon emitted from the initial state positron
of the final state anti--fermion $\bar{f}$) are to be accounted as 
the signal. These diagrams describe the non--singlet (NS) mechanism
of pair production. The diagrams of the type $(a)$ and $(b)$ with the
production of a secondary pair via a virtual $Z$-boson are treated 
as background. The so-called multi--peripheral process is described by Fig.~1$c$.
Diagrams of type $(d)$ represent the singlet mechanism of pair production.
Their contribution is considered as a background too. The pure electroweak 
four--fermion processes with $W$-boson intermediate states fall into
a separate class~\cite{Kobel:2000aw} from both the theoretical and the
experimental points of view. The relevant (signal) Feynman graphs for virtual
pair corrections are given in Figure~2.

%--------------------------Figure~2-----------------------
\begin{figure}[ht]
\begin{center}
\epsfig{file=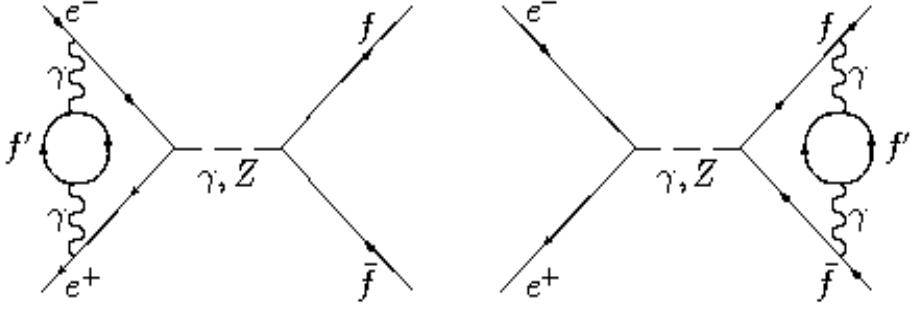,width=12cm,height=4cm}
\end{center}
\caption{Feynman diagrams for virtual pair corrections.}
\label{figure:2}
\end{figure}
%--------------------------Figure~2-----------------------

In other words we will count only the non--singlet secondary pairs produced 
via virtual photons Fig.~1$a,b$ (some numbers for the singlet pairs will
be presented ). The contribution from the so-called background diagrams, 
like multi--peripheral, double--resonant $(ZZ)$, and singlet pair production,
are supposed to be subtracted from the data. Interferences between the
{\em signal} and {\em background} amplitudes are typically small, but some
of them, like the ones between the IS singlet and non--singlet pairs,
should be taken into account. We will not use any direct cut on the phase 
space volume of real secondary pairs (an inclusive treatment of radiative 
events). Only the condition (see Eq.~(\ref{cuts})) on the primary pair invariant 
mass $s'$ will be applied (the dependence on the secondary pair invariant mass 
will be shown for the FS correction).

\subsection{The final state pair RC} 

In the lowest order for the FS pair corrections we use the results of 
Ref.~\cite{Hoang:1995ht}. Our aim here is to estimate the effect of higher
orders. Concerning the interplay with the IS photonic corrections, it is
claimed, that the final state pair corrections should be realized in the
multiplicative way:
\ba \label{multip}
\sigma = \sigma_{\mathrm{Born}}(1+\delta_{\mathrm{IS}\gamma})(1+\delta_{\mathrm{FSP}}),
\ea
where $\delta_{\mathrm{IS}\gamma}$ stands for the initial state photonic correction,
and $\delta_{\mathrm{FSP}}$ gives the final state pair one. At LEP2 energies, when the
radiative return to the $Z$-peak is allowed, we have huge values of $\delta_{\mathrm{IS}\gamma}$
(typically above 100 percent), and the additive summation of the corrections
$(1+\delta_{\mathrm{IS}\gamma}+\delta_{\mathrm{FSP}})$ fails completely. In principle, the
initial state pair correction should participate in formula~(\ref{multip}) together
with the photonic one, but, as one can see from numerical illustrations, an
additive treatment of the IS and FS pair corrections is acceptable as well.

The effect of simultaneous photon and pair radiation from the final state can be 
estimated by multiplying the corresponding contributions: $\order{\alpha}$ FS photon RC
and the $\order{\alpha^2}$ FS pair RC. One can see that this source does not give more 
than 0.05 permil in the accepted signal definition. This estimate was confirmed
also by a computation (analogous to Eq.~(\ref{sig3p})) of the $\order{\alpha^3}$
LLA result for the FS non--singlet pairs. It is worth to note here, that additional
strong cuts on the final state photons and pairs can increase the FS corrections 
and revive the problem of the corresponding missing higher order terms.

\begin{table}[ht]
\begin{tabular}[]{|c|c|c|c|c|c|c|} \hline
$E_{\mathrm{cm}}$ [GeV] $\ \backslash$ $P_{\mathrm{cut}}$ &
0.0001 & 0.001 & 0.01 & 0.1 & 0.3 & 0.99 \\ \hline
 & \multicolumn{6}{c|}{$z_{\mathrm{min}}=0.01$} \\ \hline
88.1867 & $-$0.47 & $-$0.13 & $+$0.11 & $+$0.18 & $+$0.19 & $+$0.20 \\
91.1867 & $-$0.48 & $-$0.13 & $+$0.11 & $+$0.19 & $+$0.20 & $+$0.20 \\
94.1867 & $-$0.48 & $-$0.13 & $+$0.11 & $+$0.19 & $+$0.20 & $+$0.20 \\
189     & $-$0.57 & $-$0.12 & $+$0.14 & $+$0.22 & $+$0.23 & $+$0.23 \\
206     & $-$0.59 & $-$0.11 & $+$0.14 & $+$0.22 & $+$0.23 & $+$0.23 \\
\hline
 & \multicolumn{6}{c|}{$z_{\mathrm{min}}=0.1$} \\ \hline
88.1867 & $-$0.51 & $-$0.18 & $+$0.05 & $+$0.12 & $+$0.12 & $+$0.12 \\
91.1867 & $-$0.51 & $-$0.18 & $+$0.05 & $+$0.12 & $+$0.12 & $+$0.12 \\
94.1867 & $-$0.52 & $-$0.18 & $+$0.05 & $+$0.12 & $+$0.12 & $+$0.12 \\
189     & $-$0.62 & $-$0.18 & $+$0.06 & $+$0.13 & $+$0.14 & $+$0.14 \\
206     & $-$0.64 & $-$0.18 & $+$0.06 & $+$0.13 & $+$0.14 & $+$0.14 \\
\hline
 & \multicolumn{6}{c|}{$z_{\mathrm{min}}=0.7225$} \\ \hline
88.1867 & $-$1.08 & $-$0.89 & $-$0.82 & $-$0.81 & $-$0.81 & $-$0.81 \\
91.1867 & $-$1.09 & $-$0.91 & $-$0.83 & $-$0.83 & $-$0.83 & $-$0.83 \\
94.1867 & $-$1.10 & $-$0.91 & $-$0.84 & $-$0.83 & $-$0.83 & $-$0.83 \\
189     & $-$1.38 & $-$1.14 & $-$1.05 & $-$1.05 & $-$1.05 & $-$1.05 \\
206     & $-$1.42 & $-$1.17 & $-$1.08 & $-$1.08 & $-$1.08 & $-$1.08 \\
\hline
 & \multicolumn{6}{c|}{$z_{\mathrm{min}}=0.9$} \\ \hline
88.1867 & $-$1.68 & $-$1.62 & $-$1.61 & $-$1.61 & $-$1.61 & $-$1.61 \\
91.1867 & $-$1.70 & $-$1.63 & $-$1.63 & $-$1.63 & $-$1.63 & $-$1.63 \\
94.1867 & $-$1.70 & $-$1.64 & $-$1.63 & $-$1.63 & $-$1.63 & $-$1.63 \\
189     & $-$2.13 & $-$2.04 & $-$2.04 & $-$2.04 & $-$2.04 & $-$2.04 \\
206     & $-$2.19 & $-$2.11 & $-$2.10 & $-$2.10 & $-$2.10 & $-$2.10 \\
\hline
\end{tabular}
\caption{Radiative correction $\delta^{e^+e^-\to\mu^+\mu^-}_{\mathrm{FSP}}$ in permil
for different cuts.}
\label{Table1}
\end{table}

In Table~\ref{Table1} we give the value of correction $\delta_{\mathrm{FSP}}$ due to the
final state pair production in the process of annihilation into muons as a function
on cuts on the invariant mass of the primary and secondary pairs;
\ba \label{cuts}
M^2_{\mathrm{primary}} &\equiv& s' \geq z_{\mathrm{min}} s, \qquad
M^2_{\mathrm{secondary}} \leq s P_{\mathrm{cut}}, \nonumber \\
\delta_{\mathrm{FSP}} &=& 10^3\left(\frac{\sigma^{\mathrm{RC}}_{\mathrm{ISP,FSP},\gamma}}
{\sigma^{\mathrm{RC}}_{\mathrm{ISP},\gamma}} - 1 \right),
\ea
where $\sigma^{\mathrm{RC}}_{\mathrm{ISP},\gamma}$  is the cross section with 
the photonic and IS pair corrections taken into account. One can see that the
dependence on the invariant mass of the secondary pair is very weak. 
That confirms our expectations, because the contribution of large--invariant--mass
pairs $(M^2_{\mathrm{secondary}}\gg m_f^2$, where $m_f$ is the mass of 
a pair constituent; $m_f=m_\pi$ for the case of hadronic pairs) does not
contain any large logarithm, and so it is suppressed by the relative factor $1/L^2$.
The same should be valid for the IS pairs: the relative contribution of 
large--invariant--mass secondary pairs is negligible. This statement concerns,
of course, pair production via virtual photons, while the double--resonant
$ZZ$-mechanism (or $WW$) has to be treated separately.

\subsection{The initial state pair RC} 

The combination of the initial state pair correction with the final state 
photonic one should be treated in the same way as Eq~(\ref{multip}), while
it is not so important numerically. We will concentrate now on the main
problem, when both the pair and the photons are produced in the initial
state.

The initial state pair contributions to the corrected cross section
is presented as the convolution of the {\em boosted} Born cross section
with the so-called pair radiator:
\ba \label{integ}
\delta\sigma^{\mathrm{pair}} = \int^1_{z_{\mathrm{min}}}\dd z\,
\tilde{\sigma}(zs) H(z) = \sigma(s) H_\Delta
+ \int^{1-\Delta}_{z_{\mathrm{min}}}\dd z\, \tilde{\sigma}(zs) H_\Theta(z).
\ea 
Where $H_\Delta$ represents the impact of virtual and soft (small--energy) 
pairs, $\Delta$ is an auxiliary parameter $(m_e^2/s \ll \Delta \ll 1$);
numerical results should be independent of its value. In higher orders,
when we have instantaneous emission of pairs and photons, we will still 
call such corrections as the pair ones. In our computations the
$\order{\alpha^2}$ formula for the pair radiator is taken from 
Refs.~\cite{Berends:1987ab,Kniehl:1988id}.

Going beyond the lowest order, one starts usually with the leading logarithmic
approximation. Indeed, the main contribution to the radiative correction
under consideration is coming from the electron pair production, because it
is reinforced by powers of the large logarithm $L_e\equiv\ln(s/m_e^2)$
(the corresponding leading logs for pairs of other flavours can be constructed 
as well). There are two different mechanisms for electron--positron pair creation
in our process: the singlet and non--singlet ones. The first one is absent for
non--electron pairs.

At the $Z$-peak it is worth to try to re--sum higher orders, because the convergence
of the perturbation theory there is slow. The first exponentiated formula for the
IS LLA pair production was suggested in Ref.~\cite{Kuraev:1985hb}. The process
of one pair production was supplied there by emission of an arbitrary number
of photons. This approach gives a reasonable approximation for the leading 
corrections close to the $Z$-peak. But it does not include some significant
next--to--leading terms, and even the known third order leading logs are not
reproduced completely.

Another way~\cite{Catani:1989et} is to use the Mellin transformation to solve 
the QED evolution equation for soft and collinear photonic contributions
supplied by the running QED coupling constant. The authors received a
simultaneous exponentiation of photonic and pair corrections, which takes
into account important next--to--leading terms. Unfortunately, their formulae 
require additional evaluations (inverse Mellin transformations). Moreover, 
the applicability of their method to the mixed pair--photon corrections
is not clear.

In Ref.~\cite{Jadach:1992aa} a phenomenological formula for simultaneous 
exponentiation of photonic and pair radiation was proposed. The correspondence 
of the exponentiated formula to the perturbative results was shown for the case 
of real hard radiation. Nevertheless, the structure of the radiator function, 
suggested in \cite{Jadach:1992aa}, does not allow to verify analytically the 
correspondence for the soft and virtual part of the corrections.

\subsubsection{The leading logarithmic approximation}

Let us start our evaluations with the leading logarithmic approximation. 
It will serve us as a guide. The third order LLA radiator function can be
taken from the Dokshitzer--Gribov--Lipatov--Altarelli--Parisi
evolution equation (we dropped below the pure photonic corrections):
\ba \label{sig3p}
\delta\sigma_{e}^{(3){\mathrm{pair}}} = \int\dd z\, \tilde\sigma(zs) 
\biggl(\frac{\alpha}{\pi}(L_e-1)\biggr)^3
\biggl[ \frac{5}{24}P^{(1)}\otimes R^{s}(z)
- \frac{1}{36}R^{s}(z) + \frac{1}{6}P^{(2)}(z) 
+ \frac{1}{27}P^{(1)}(z) \biggr],
\ea
where
\ba \label{Rs}
R^{s}(z) = \frac{1-z}{3z}(4+7z+4z^2) + 2(1+z)\ln z
\ea
is the singlet splitting function;
\begin{eqnarray}
P^{(1)}(z) &=&  \lim_{\Delta\to 0}\biggl\{ \delta(1-z)\biggl(
2\ln\Delta + \frac{3}{2} \biggr) +  \Theta(1-\Delta -z)
\frac{1+z^2}{1-z} \biggr\},  \\ \nonumber
P^{(2)}(z) &=& \lim_{\Delta\to 0}\biggl\{ \delta(1-z)\biggl[
\biggl(2\ln\Delta + \frac{3}{2}\biggr)^2 - \frac{2\pi^2}{3} \biggr]
\nonumber \\
&+& \Theta(1-\Delta -z)2\biggl[ \frac{1+z^2}{1-z}\biggl(
2\ln(1-z)-\ln z + \frac{3}{2}\biggr) + \frac{1+z}{2}\ln z - 1 + z
\biggr] \biggr\}
\end{eqnarray}
are the non--singlet splitting functions.
The convolution of the singlet splitting function $R^{s}$
with the first order non--singlet one is known~\cite{Skrzypek:1992vk}:
\begin{eqnarray}
P^{(1)}\otimes R^{s}(z) &=& \biggl( \frac{3}{2} + 2\ln(1-z) \biggr)
R^{s}(z)
+ (1+z)( - \ln^2z + 4\Li(1-z) ) \nonumber \\
&+& \frac{1}{3}( - 9 - 3z + 8z^2 )\ln z
+ \frac{2}{3}\biggl( - \frac{3}{z} - 8 + 8z + 3z^2 \biggr).
\end{eqnarray}

\subsubsection{Beyond the LLA}

But, as we learned from the $\order{\alpha^2}$ level from the complete
calculations, the contribution of the sub--leading correction in the
region close to the $Z$-peak is very large (say, about one half of the 
leading one). In order to account the most significant part of the 
sub--leading corrections we consider the
convolution of the ${\mathcal O}(\alpha^2)$  pair radiator with
the ordinary ${\mathcal O}(\alpha)$ photonic radiator, proportional
to the $P^{(1)}$ splitting function. In this way we receive 
the main part of the ${\mathcal O}(\alpha^3)$ leading logs,
proportional to $P^{(2)}$, and the sub--leading terms enhanced 
by $\ln(1-z)/(1-z)$, like $L^2\ln(1-z)/(1-z)$ and $L\ln^2(1-z)/(1-z)$.
Note that the convolution as well as exponentiation can not
give the true complete sub--leading formula. And, contrary 
to the case of pure photonic radiation, they are unable even
to cover the full leading log result: the term with $P^{(1)}$
is not reproduced (this term however has a small coefficient and no
any $\ln(1-z)$, so it is negligible numerically in our task). 
This term corresponds to the physical situation (Feynman diagrams),
when the extra photon is attached to the secondary pair.

In fact the convolution 
gives a part of sub-leading terms coming from the kinematical regions,
where both the pair and the photon are either soft or emitted collinearly
to the parent particle momentum, while there 
are other sources for the corrections, like, for instance, 
emission of a collinear pair and a large--angle photon.
But we believe, that the main terms with reinforcements are reproduced
correctly: that follows from the general experience in leading
log calculations. Moreover, for the case of pure photonic radiation 
this has been checked by direct perturbative calculations.

In this way we received the expressions for the third order pair radiator.
For leptonic non--singlet pairs we have
\ba \label{hard3f}
H^{(3)f}_{\Theta}(z) &=& \biggl(\frac{\alpha}{\pi}\biggr)^3L_e\biggl[
\frac{1+z^2}{1-z}\, L_f^2\biggl(\frac{2}{3}\ln(1-z) - \frac{1}{3}\ln z
+ \frac{1}{2} \biggr)
% \nonumber \\ &+& 
+ L_f^2\biggl( \frac{1+z}{6}\ln z - \frac{1-z}{3} \biggr) \nonumber \\
&+& \frac{1+z^2}{1-z}\, L_f\biggl( 2\ln^2(1-z) - \frac{11}{9}\ln(1-z)
- \frac{9}{4} - \frac{4}{3}\zeta(2) - 2\ln z\ln(1-z) \nonumber \\ 
&+& \frac{1}{3}\ln^2z
+ \frac{11}{18}\ln z \biggr) 
+L_f\biggl( - \frac{8}{3}(1-z)\ln(1-z) + \frac{2}{3}(1+z)\ln z \ln(1-z)
\nonumber \\
&-& \frac{1}{6}(1+z)\ln^2z + \frac{4}{9}(1-5z)\ln z 
+ \frac{2}{3}(1+z)\Li(1-z) + \frac{19}{9}(1-z) \biggr) \nonumber \\
&+& \frac{1+z^2}{1-z}\biggl( \frac{16}{9}\ln^3(1-z) 
- \frac{7}{3}\ln^2(1-z) + \frac{67}{27}\ln(1-z) 
- \frac{16}{3}\zeta(2)\ln(1-z) \nonumber \\
&-& \frac{8}{3}\ln z\ln^2(1-z) + \frac{7}{3}\ln z\ln(1-z) 
+ \frac{5}{6}\ln^2z\ln(1-z) - \frac{1}{3}\Li(1-z)\ln(1-z) \nonumber \\
&-&  \frac{1}{18}\ln^3z - \frac{31}{72}\ln^2z - \frac{67}{54}\ln z 
- \frac{2}{3}\Li(1-z)\ln z + \frac{8}{3}\zeta(2)\ln z 
- \frac{1}{4}\Li(1-z)  \nonumber \\
&-& \frac{5}{3}{\mathrm S}_{1,2}(1-z) - \frac{4}{3}\zeta(2)
+ 4\zeta(3) + \frac{1073}{162}  \biggr)
+ {\mathcal O}(1) \biggr], \\ \nonumber 
\Li(x) &=& -\int\limits_{0}^{x}\frac{\dd y}{y}\ln(1-y), \qquad
{\mathrm S}_{1,2}(x) = \frac{1}{2}\int\limits_{0}^{x}
\frac{\dd y}{y}\ln^2(1-y).
\ea
The large logarithm $L_f=\ln(s/m_f^2)$ stands for different
lepton flavours: $f=e$, $\mu$, $\tau$.
The corresponding {\em soft+virtual} contribution is given by
\begin{eqnarray} \label{sv3f}
H^{(3)f}_{\Delta} &=& 
\biggl(\frac{\alpha}{\pi}\biggr)^3L_e\biggl[
\frac{8}{9}\ln^4\Delta + \ln^3\Delta\biggl(\frac{4}{3}L_f 
- \frac{14}{9}\biggr) 
+ \ln^2\Delta\biggl( \frac{2}{3}L_f^2 - \frac{11}{9}L_f 
+ \frac{67}{27} - \frac{16}{3}\zeta(2)\biggr)
\nonumber \\
&+& \ln\Delta\biggl( L^2_f - \frac{9}{2}L_f
- \frac{8}{3}\zeta(2)L_f
+ 8\zeta(3) - \frac{8}{3}\zeta(2) + \frac{1073}{81} \biggr) 
+ L_f^2\left(\frac{3}{8}-\frac{2}{3}\zeta(2)\right)
\nonumber \\
&+& L_f\left(- \frac{17}{8} + \frac{8}{3}\zeta(3)+\frac{20}{9}\zeta(2)\right)
+ \frac{1543}{216} - \frac{22}{9}\zeta(3)
- \frac{413}{54}\zeta(2) + \frac{8}{15}\zeta^2(2) \biggr].
\end{eqnarray}
The cancellation of the auxiliary parameter $\Delta$ was checked
analytically and achieved numerically. The largest contribution 
appears at the $Z$-peak,
where the integral over $z$ is practically limited by the $Z$-boson
width: $(z_{\mathrm{min}},1) \to (1-\Gamma_Z/M_Z,1)$. And therefore
integrals from terms, like $\ln(1-z)/(1-z)$, are large.
We checked that for the real hard emission there is an agreement
between the most important terms in $H^{(3)}_{\Theta}(z)$ and
the corresponding terms in expansion of the exponentiated formula
from Ref.~\cite{Jadach:1992aa}. Such a correspondence between the exponentiation
and convolution procedures is well known also 
in the case of pure photonic radiation.

A similar expression was derived for the case of hadronic pairs:
\begin{eqnarray}  \label{h3ht}
H^{(3)\mathrm{h}}_{\Theta}(z) &=& 
\biggl(\frac{\alpha}{\pi}\biggr)^3L_e\biggl\{
\frac{1+z^2}{1-z}\, L_h^2\Ri\biggl(\frac{2}{3}\ln(1-z) 
- \frac{1}{3}\ln z + \frac{1}{2} \biggr)  \nonumber \\
&+& L_h^2\Ri\biggl( \frac{1+z}{6}\ln z - \frac{1-z}{3} \biggr)
+ \frac{1+z^2}{1-z}\, L_h\biggl[ \Ri\biggl( 2\ln^2(1-z) 
+ \ln(1-z)   \nonumber \\
&-& \frac{8}{3}\ln2\ln(1-z) - 2\ln(z)\ln(1-z)
+ \frac{1}{3}\ln^2z - \frac{1}{2}\ln z + \frac{4}{3}\ln2\ln z
- \frac{7}{12} - 2\ln2  \nonumber \\ 
&-& \frac{4}{3}\zeta(2) \biggr) 
+ R_0\biggl( \frac{4}{3}\ln(1-z) + 1 
- \frac{2}{3}\ln z \biggr) \biggr]
+ L_h\biggl[ \Ri\biggl( - \frac{8}{3}(1-z)\ln(1-z)  \nonumber \\
&+& \frac{2}{3}(1+z)\ln(z)\ln(1-z) - \frac{1}{6}(1+z)\ln^2(z) 
+ \frac{2}{3}(1+z)\Li(1-z) + \ln z  \nonumber \\
&-& \frac{5}{3}z\ln z + \frac{2}{3}\ln2(1+z)\ln z
 + \frac{4}{3}(1-z)\ln2 + 1 - z
\biggr) \nonumber \\
&+& R_0\biggl( \frac{1}{3}(1+z)\ln z - \frac{2}{3}(1-z) \biggr)
\biggr] 
+ \frac{1+z^2}{1-z}\biggl[ \Ri\biggl( \frac{16}{9}\ln^3(1-z)
+ \ln^2(1-z) \nonumber \\
&-& 4\ln2\ln^2(1-z) 
- 4\zeta(2)\ln(1-z) 
- 2\ln2\ln(1-z)
+ \frac{8}{3}\ln^22\ln(1-z)  \nonumber \\
&-& \frac{1}{3}\Li(1-z)\ln(1-z)
+ \frac{5}{6}\ln^2z\ln(1-z)
- \frac{8}{3}\ln z\ln^2(1-z) \nonumber \\
&-& \ln z\ln(1-z)
+ 4\ln2\ln z\ln(1-z)
- \frac{5}{3}\mathrm{S}_{1,2}(1-z)
- \frac{2}{3}\Li(1-z)\ln z \nonumber \\
&-& \frac{1}{4}\Li(1-z)
- \frac{1}{18}\ln^3z
+ \ln2\ln z
+ 2\zeta(2)\ln z
+ \frac{1}{8}\ln^2z
- \frac{2}{3}\ln2\ln^2z  \nonumber \\
&-& \frac{4}{3}\ln^22\ln z
+ \frac{8}{3}\ln2\zeta(2)
+ \frac{7}{6}\ln2
+ 2\ln^22 - \zeta(2) 
+ \frac{10}{3}\zeta(3) 
+ \frac{5}{8}
\biggr)  \nonumber \\
&+& R_0\biggl( 2\ln^2(1-z) - \frac{8}{3}\ln2\ln(1-z) + \ln(1-z)
- 2\ln z\ln(1-z) \nonumber \\
&-& \frac{1}{2}\ln z + \frac{1}{3}\ln^2z 
+ \frac{4}{3}\ln2\ln z - \frac{7}{12} - 2\ln2 
- \frac{4}{3}\zeta(2) \biggr)\nonumber \\
&+& R_1\biggl( \frac{4}{3}\ln(1-z) + 1 - \frac{2}{3}\ln z  \biggr)
\biggr] + {\mathcal O}(1) \biggr\}, 
\end{eqnarray}
\begin{eqnarray}  \label{h3hsv}
H^{(3)\mathrm{h}}_{\Delta} &=& 
\biggl(\frac{\alpha}{\pi}\biggr)^3L_e\biggl\{
\ln^4\Delta\frac{8}{9}\Ri 
+ \ln^3\Delta\biggl[ \Ri\biggl( \frac{4}{3}L_h + \frac{2}{3}
- \frac{8}{3}\ln2\biggr) + \frac{4}{3}R_0 \biggr] \nonumber \\
&+& \ln^2\Delta\biggl[ \Ri\biggl( \frac{2}{3}L^2_h + L_h -
\frac{8}{3}\ln2L_h - 2\ln2 + \frac{8}{3}\ln^22 - 4\zeta(2) \biggr) 
\nonumber \\
&+& R_0\biggl( \frac{4}{3}L_h + 1 - \frac{8}{3}\ln2 \biggr) 
+ \frac{4}{3}R_1 \biggr]
+ \ln\Delta\biggl[ \Ri\biggl( L_h^2 - \frac{8}{3}\zeta(2)L_h
\nonumber \\
&-& \frac{7}{6}L_h - 4\ln2L_h + \frac{16}{3}\ln2\zeta(2) 
+ \frac{7}{3}\ln2 + 4\ln^22
- 2\zeta(2) + \frac{20}{3}\zeta(3) \nonumber \\
&+& \frac{5}{4} \biggr)
+ R_0\biggl( 2L_h - \frac{8}{3}\zeta(2) - \frac{7}{6} - 4\ln2\biggr)
+ 2R_1 \biggr] \nonumber \\  
&+& \left(\frac{3}{8}-\frac{2}{3}\zeta(2)\right)L_h^2\Ri 
+ L_h\Ri\biggl(\frac{8}{3}\zeta(3) + \frac{8}{3}\ln 2 \zeta(2) - \frac{7}{8} 
- \frac{3}{2}\ln2 \biggr)\nonumber \\  
&+& L_hR_0\biggl(\frac{3}{4} - \frac{4}{3}\zeta(2)\biggr)
+ R(\infty)\biggl(\frac{3}{2}\ln^22 + \frac{7}{4}\ln2 + \zeta(3)
- \frac{16}{3}\zeta(3)\ln2 - \frac{4}{5}\zeta^2(2) \nonumber \\  
&-& \frac{5}{12}\zeta(2)
- \frac{8}{3}\zeta(2)\ln^22 + \frac{23}{48} \biggr)
+ R_0\biggl(\frac{8}{3}\zeta(3) + \frac{8}{3}\zeta(2)\ln2 - \frac{3}{2}\ln2 
- \frac{7}{8}\biggr)  \nonumber \\  
&+& R_1\biggl(\frac{3}{4}-\frac{4}{3}\zeta(2)\biggr) \biggr\}, 
\qquad L_h = \ln\frac{s}{m^2_\pi}\, .
\end{eqnarray}
Quantities $R(\infty)$, $R_{0,1}$ are defined in Ref.~\cite{Kniehl:1988id}.
In contrast to Ref.~\cite{Jadach:1992aa} we extended the hadronic
pair contribution to the third order {\em dynamically}, but
not by a static coefficient. That is really important at LEP2 energies.

The LLA leptonic contributions, which were not
reproduced by the convolution (the terms with $P^{(1)}$ 
and $P^{(1)}\otimes R^s$ functions in Eq.~(\ref{sig3p})),
should be added to Eqs.~(\ref{hard3f}), (\ref{sv3f}).
The third order effect appeared to be not small numerically.
So, we have to look also at the fourth order. 
The corresponding contribution is
estimated by means of the leading logs~\cite{Arbuzov:1999cq} 
(non--singlet channel only):
\begin{eqnarray} \label{quadr}
\dd\sigma_{e}^{(4)} = \int\dd z \tilde\sigma(zs) 
\biggl(\frac{\alpha}{2\pi}(L_e-1)\biggr)^4
\biggl[ 
\frac{1}{12}P^{(3)}(z)
+ \frac{11}{216}P^{(2)}(z) 
+ \frac{1}{108}P^{(1)}(z) \biggr].
\end{eqnarray}
Only the term with the $P^{(3)}$ splitting
function~\cite{Skrzypek:1992vk} seems to be numerically significant. 
It agrees with the corresponding
term in the expansion of exponentiated formula in Ref.~\cite{Jadach:1992aa}.
The size of the fourth order, which is estimated from the above
formula for electron pairs, is rather small.
So, we can be less rigorous: in order to estimate the fourth order
contribution from other kinds of pairs and the large sub--leading
terms, we drop them simultaneously. That was checked in lower orders
to be a reasonable approximation. So, in the ${\mathcal O}(\alpha^4)$ 
we keep only the leading logarithmic formula~(\ref{quadr}) 
for non--singlet electron pairs.

\begin{table}[ht]
\begin{tabular}[]{|c|c|c|c|c|c|c|c|} \hline
 & $e$(NS) & $e$(S) & $e$(NS$\times$S) & $\mu$ & $\tau$ & hadr. & sum(NS) \\ \hline
 & \multicolumn{7}{c|}{$E_{\mathrm{cm}}=88.1867$~GeV} \\ \hline
${\mathcal O}(\alpha^2)$ & 
$-$2.11 & $+$1.13 & $-$0.02 & $-$0.34 & $-$0.04 & $-$0.70 & $-$3.19 \\ 
${\mathcal O}(\alpha^3)$ & 
$-$1.51 & $+$1.12 & $-$0.02 & $-$0.26 & $-$0.04 & $-$0.35 & $-$2.16 \\ 
${\mathcal O}(\alpha^4)$ & 
$-$1.70 & $+$1.12 & $-$0.02 & $-$0.26 & $-$0.04 & $-$0.35 & $-$2.36 \\ \hline
 & \multicolumn{7}{c|}{$E_{\mathrm{cm}}=91.1867$~GeV} \\ \hline
${\mathcal O}(\alpha^2)$ & 
$-$2.25 & $+$0.16 & $-$0.01 & $-$0.38 & $-$0.03 & $-$0.77 & $-$3.43 \\ 
${\mathcal O}(\alpha^3)$ & 
$-$1.69 & $+$0.15 & $-$0.01 & $-$0.30 & $-$0.04 & $-$0.42 & $-$2.44 \\ 
${\mathcal O}(\alpha^4)$ & 
$-$1.78 & $+$0.15 & $-$0.01 & $-$0.30 & $-$0.04 & $-$0.42 & $-$2.54 \\ \hline
 & \multicolumn{7}{c|}{$E_{\mathrm{cm}}=94.1867$~GeV} \\ \hline
${\mathcal O}(\alpha^2)$ & 
$+$1.25 & $+$0.55 & $-$0.05 & $+$0.09 & $-$0.02 & $+$0.05 & $+$1.37 \\ 
${\mathcal O}(\alpha^3)$ & 
$+$0.13 & $+$0.54 & $-$0.05 & $-$0.10 & $-$0.03 & $-$0.21 & $-$0.22 \\ 
${\mathcal O}(\alpha^4)$ & 
$+$0.43 & $+$0.54 & $-$0.05 & $-$0.10 & $-$0.03 & $-$0.21 & $+$0.08 \\ \hline
 & \multicolumn{7}{c|}{$E_{\mathrm{cm}}=189$~GeV} \\ \hline
${\mathcal O}(\alpha^2)$ & 
$+$6.33 & $+$31.80 & $-$1.32 & $+$1.93 & $+$0.63 & $+$5.30 & $+$14.20\\ 
${\mathcal O}(\alpha^3)$ & 
$+$7.13 & $+$30.48 & $-$1.32 & $+$2.11 & $+$0.68 & $+$5.83 & $+$15.76\\ 
${\mathcal O}(\alpha^4)$ & 
$+$7.08 & $+$30.48 & $-$1.32 & $+$2.11 & $+$0.68 & $+$5.83 & $+$15.71\\ \hline
 & \multicolumn{7}{c|}{$E_{\mathrm{cm}}=206$~GeV} \\ \hline
${\mathcal O}(\alpha^2)$ & 
$+$6.43 & $+$40.38 & $-$1.41 & $+$2.00 & $+$0.68 & $+$5.56 & $+$14.68\\ 
${\mathcal O}(\alpha^3)$ & 
$+$7.32 & $+$38.74 & $-$1.40 & $+$2.22 & $+$0.74 & $+$6.19 & $+$16.48\\ 
${\mathcal O}(\alpha^4)$ & 
$+$7.28 & $+$38.74 & $-$1.40 & $+$2.22 & $+$0.74 & $+$6.19 & $+$16.44\\ \hline
\end{tabular}
\caption{Different contributions to $\delta_{\mathrm{ISP}}^{e^+e^-\to \mathrm{hadr.}}$
in permil.}
\label{Table2}
\end{table}

In Table~\ref{Table2} we present the results for the different contributions
of pair corrections to the process of annihilation into hadrons. 
The value of correction due to pairs is defined in respect to
the cross section for annihilation into hadrons with pure
photonic corrections taken into account:
\ba
\delta_{\mathrm{ISP}} = 10^3\left(\frac{\sigma^{\mathrm{RC}}_{\mathrm{ISP},\gamma}}
{\sigma^{\mathrm{RC}}_{\gamma}} - 1\right).
\ea 
The cut-off on the primary pair mass $s' > z_{\mathrm{min}}\cdot s$,
$\ \ z_{\mathrm{min}}=0.01$, acts in the same way on real photons and secondary pairs.

In Table~\ref{Table3} we present the values of $\delta^{e^+e^-\to\mu^+\mu^-}_{\mathrm{ISP}}$
for different cuts and beam energies. In the first two columns we give the contributions of
singlet pairs and their interference with the non--singlet ones. The latter, in principle,
should be counted as a part of the signal. The third column contains the NS contribution
(our {\em best estimate} with all pair flavours summed up). It can be compared with our 
implementation of exponentiated pair corrections. Note, that in both the last two columns
the contribution of hadronic pairs is not exponentiated, but it is supplied according to
our approach (we used result~\cite{Kniehl:1988id} supplied with our third order contribution
(\ref{h3ht}), (\ref{h3hsv})).

For the column ``\cite{Kuraev:1985hb}'', the known complete second order
result was supplied by the NS higher order part of the exponentiated formula, 
derived in this paper.  The discrepancy with our {\em best estimate} 
originates from an improper reproduction of the third order correction
by the exponentiation in Ref.~\cite{Kuraev:1985hb}. Note that the largest
difference appears at LEP2 energies for small $z_{\mathrm{min}}$ values,
which means the dominance of the effect of radiative return to the
$Z$-peak due to hard pair emission. While the exponentiation~\cite{Kuraev:1985hb}
is well justified only for soft radiation, which was the only one relevant at LEP1.
The last column was calculated by using the exponentiated formula 
of Ref.~\cite{Jadach:1992aa}. The approximation for the higher order 
hadronic contribution (static coefficient before the second order contribution),
suggested in that paper has sense only at LEP1 energy on the top
of the peak and the left wing; everywhere else it fails. Errors in the
third column represent our estimate of theoretical uncertainties
(see the Conclusions).

As could be seen from Table~\ref{Table3}, the contribution of singlet pair
production becomes significant for small values of $z_{\mathrm{min}}$.
In data analysis such events are supposed~\cite{Kobel:2000aw} to be
extracted from the data together with the two--photon (multi--peripheral)
process~\cite{Decamp:1990ky}. We underline, that the procedure should 
be accurate and well understood, because the events with singlet pairs 
and multi--peripheral production have completely different kinematics
and signatures in the detector. At LEP2 energies the contribution
of singlet pairs becomes really important, if the returning to the
$Z$-peak is allowed (for $z_{\mathrm{min}} \la 0.25$). The large contribution
is coming from the peculiar $1/z$ term in the function $R^s$, Eq.~(\ref{Rs}).
Note, that this function is
multiplied by the kernel cross section which contains
$z^{-1}$ too. In this way the second order correction due to 
collinear emission of singlet pairs can give several percent.  
This did not happen at LEP1, because the kernel cross section
outside the $Z$-peak is relatively small.

The difference between the column ``non--singlet'' and the last
one ``\cite{Jadach:1992aa}$^{*}$'' is coming just from the difference
of the perturbative and exponentiated treatment of higher
order leptonic pair corrections. The difference is less
on the right wing of the $Z$-peak, where the real radiation
is dominant, because it provides returning to the peak.
For real radiation we expanded the exponentiated formula
and saw an agreement in the main terms. On the other side
of the peak (and on the top) the virtual and soft corrections
are dominant, because any real hard radiation leads to a
huge reduction of the kernel cross section in Eq.~(\ref{integ}).
In this case we have to compare the {\em soft+virtual} part of our
radiator $H_{\Delta}$ with the integral 
\begin{equation} \label{jmssv}
\int\limits_{1-\Delta}^{1}\dd z\, (\rho(z)-\rho^{\gamma}(z)),
\end{equation}
where $\rho(z)$ is the exponentiated formula~\cite{Jadach:1992aa}, and
$\rho^{\gamma}(z)$ is the pure photonic exponentiated 
radiator~\cite{Jadach:1990vz}.
The latter is in perfect agreement
with the perturbative results both in the hard and soft+virtual
regions, while the formula with simultaneous exponentiations
of photonic and pair radiation has a very peculiar structure
for $z\to 1$, and so Eq.~(\ref{jmssv}) can be hardly expanded in a 
series in $\alpha$ to compare with the perturbative results.  
The problem is seen
in the region of very soft photons $(z\to 1)$ below 
the threshold of pair production, where the part of the
exponentiated formula, which describes pairs,
still represents a sharp dependence on $z$-value,
instead of giving a constant coefficient from virtual pairs.

It is worth to note the weak dependence on the $z_{\mathrm{min}}$
parameter of the corrections at LEP1 energies. 
So, the bulk of the corrections
is given by soft and virtual pairs, which are not mixed during 
the observation with the primary pair from electron positron
annihilation.

\begin{table}[ht]
\begin{tabular}[]{|c|c|c|c|c|c|} \hline
$E_{\mathrm{cm}}$ [GeV] & singlet & S$\times$NS & non--singlet
 & \cite{Kuraev:1985hb} & \cite{Jadach:1992aa}$^{*}$  
\\ \hline
 & \multicolumn{5}{c|}{$z_{\mathrm{min}}=0.01$} \\ \hline
88.1867 & $+$5.61 & $-$0.06 & $-$1.96~~$\pm$0.28 & $-$1.52 & $-$1.76 \\
91.1867 & $+$0.81 & $-$0.01 & $-$2.48~~$\pm$0.22 & $-$2.16 & $-$2.31 \\
94.1867 & $+$2.78 & $-$0.07 & $+$0.22~~$\pm$0.41 & $+$0.31 & $+$0.16 \\
189     & $+$51.03& $-$1.08 & $+$11.78 $\pm$0.82 & $+$14.17& $+$11.39\\
206     & $+$56.69& $-$1.13 & $+$12.17 $\pm$0.88 & $+$14.77& $+$11.74\\ \hline
 & \multicolumn{5}{c|}{$z_{\mathrm{min}}=0.1$} \\ \hline
88.1867 & $+$0.24 & $-$0.03 & $-$2.17~~$\pm$0.28 & $-$1.78 & $-$1.94 \\
91.1867 & $+$0.03 & $-$0.01 & $-$2.51~~$\pm$0.22 & $-$2.20 & $-$2.33 \\
94.1867 & $+$0.12 & $-$0.05 & $+$0.12~~$\pm$0.41 & $+$0.19 & $+$0.07 \\
189     & $+$12.48& $-$0.91 & $+$10.80 $\pm$0.81 & $+$12.81& $+$10.53\\
206     & $+$18.14& $-$0.95 & $+$11.13 $\pm$0.86 & $+$13.40& $+$10.83\\ \hline 
 & \multicolumn{5}{c|}{$z_{\mathrm{min}}=0.5$} \\ \hline
88.1867 & $+$0.00 & $-$0.01 & $-$2.41~~$\pm$0.28 & $-$2.04 & $-$2.17 \\
91.1867 & $+$0.00 & $-$0.01 & $-$2.54~~$\pm$0.22 & $-$2.23 & $-$2.36 \\
94.1867 & $+$0.00 & $-$0.04 & $-$0.00~~$\pm$0.41 & $+$0.06 & $-$0.05 \\
189     & $+$0.02 & $-$0.05 & $-$1.05~~$\pm$0.09 & $-$0.80 & $-$0.96 \\
206     & $+$0.02 & $-$0.05 & $-$1.09~~$\pm$0.08 & $-$0.83 & $-$1.00 \\ \hline
 & \multicolumn{5}{c|}{$z_{\mathrm{min}}=0.9$} \\ \hline
88.1867 & $+$0.00 & $-$0.01 & $-$2.58~~$\pm$0.29 & $-$2.20 & $-$2.33 \\
91.1867 & $+$0.00 & $-$0.01 & $-$2.59~~$\pm$0.22 & $-$2.28 & $-$2.41 \\
94.1867 & $+$0.00 & $-$0.04 & $-$0.32~~$\pm$0.41 & $-$0.25 & $-$0.37 \\
189     & $+$0.00 & $-$0.02 & $-$2.33~~$\pm$0.15 & $-$2.03 & $-$2.22 \\ 
206     & $+$0.00 & $-$0.02 & $-$2.40~~$\pm$0.16 & $-$2.10 & $-$2.29 \\ \hline
\end{tabular}
\caption{Radiative correction  $\delta$ for different cuts.}
\label{Table3}
\end{table}

\section{Conclusions}

First we have to remind about the effect of the initial--final 
interference (IFI) in pair corrections. We suggest to estimate it 
by using the known effect for the IFI in photonic corrections.
The bulk of the IFI pair correction is provided by the terms of
the order $\order{\alpha^2L}$ (the sub--leading logarithms). 
The terms are coming from the kinematical region of 
small--invariant--mass real pairs (and from the corresponding region
of integration over virtual pairs). Thus, one can use the
factorization of the correction factors, corresponding to the
photon emission and the pair production. So, we just multiply 
the IFI photonic correction by the factor $\alpha L/(3\pi)$ and
receive the estimate for the IFI pair RC $(L\equiv \ln(s/m^2_{e,\mu,\pi}))$.
The typical value of the photonic IFI RC at LEP energies is a few permil,
therefore one can safely neglect the corresponding pair contribution.
Note, that in this way we considered only the non--singlet mechanism
of pair production. The interference of the initial state singlet pair
production amplitudes with the final state NS ones is expected to be of 
the same magnitude. 

Let us discuss now the uncertainty in the pair corrections.
We mean just the indefiniteness in the treatment of hadronic pairs 
and of the $\order{\alpha^3}$ (and higher order) leptonic pair corrections.
Other sources, relevant for concrete experiments, like the problem
to distinguish between the primary and secondary pair or the
proper subtraction of the background, have to be discussed 
elsewhere~\cite{Kobel:2000aw}.

First we suppose, that 15\% is a reasonably safe estimate for the error 
in the second order hadronic pair contribution itself. In this case we 
use a phenomenological approach based on the measurements of 
electron--positron annihilation into hadrons at low energies.
The absence on the market of a {\em plug--and--play} parameterization
of the $R$-function (ratio of the annihilation cross sections
into hadrons and muons) does not allow to put a better precision 
level. For our computations of the FS hadronic pairs a simple
parameterization of the function was constructed. The dispersion
relations were used to look at the corresponding vacuum polarization effect
and check the function. The stability in respect of the fluctuations 
in involved data was verified. A good agreement was observed with 
the published numbers on hadronic pair 
corrections~\cite{Kniehl:1988id,Hoang:1995ht} in both the initial 
and the final states.

The uncertainty of the higher order leptonic pairs is derived as follows.
We assume that the leading logs $L$, especially the ones multiplied by a
power of $\ln(M^2_Z/\Gamma_Z^2)$, are taken into account completely.
As stated above, all sub-leading terms can not be generated neither
by convolution nor by exponentiation. Therefore, we can estimate
the error of the third order contribution to be about $1/L$
(actually we take $1/5$) of its own size. For the third order hadronic
contribution, derived by means of convolution, we should take 
into account also the intrinsic uncertainty in hadronic pairs, 
discussed above. As concerning the fourth and higher orders, we suggest
to take the error just of the order $\order{\alpha^4L^4}$ $e^+e^-$
contribution (for all flavours together).

The situation is better for the case of the final state pair corrections:
the phenomenological treatment of hadronic pairs and the omitted effect 
of the interplay between the FS photons and pairs can not lead to more 
than 10 percent error in the FS pair RC, which is already small itself.
Remind also the interference between the initial and final pair
production amplitudes, discussed above. Collecting the sources of
theoretical uncertainties, involved into our approach, we conclude:
{\em i)} at LEP1 energies the error is about 0.4 permil on the wings
and 0.2 permil on the top of the resonance; {\em ii)} at LEP2, when 
the radiative return to the $Z$-peak is forbidden, we have a reasonably
good error of about 0.1 permil (for the realistic cut 
$z_{\mathrm{min}}=0.7225)$; {\em iii)} when the radiative 
return is allowed, the size of the error can reach 1 permil.

The approach to pair corrections described here is implemented
into the semi--analytical code {\tt ZFITTER v.6.30}~\cite{Bardin:1999yd}.
A treatment of higher order pairs (with pair--photon interplay)
is available also in {\tt TOPAZ0}~\cite{Montagna:1998kp}, 
{\tt GEANTLE}~\cite{Bardin:1997zz},
{\tt KORALW}~\cite{Jadach:1998gi}, and some other codes.
Some comparisons of the approaches were done in Ref.~\cite{Kobel:2000aw}.

\subsection*{Acknowledgments}

I am grateful to D.~Bardin and G.~Passarino for valuable discussions.

\end{document}